\documentclass{moriond}

\usepackage{cite}
\newcommand*\aap{A\&A}

\newcommand*\apj{ApJ}
\newcommand*\apjl{ApJ}

\newcommand*\apjs{ApJS}

\newcommand*\araa{ARA\&A}

\newcommand*\mnras{MNRAS}

\newcommand*\pasp{PASP}

\newcommand*\prd{Phys.~Rev.~D}

\def\be{\begin{equation}}
\def\ee{\end{equation}}
\def\bea{\begin{eqnarray}}
\def\eea{\end{eqnarray}}

\newcommand{\Photo}{\includegraphics[height=35mm]{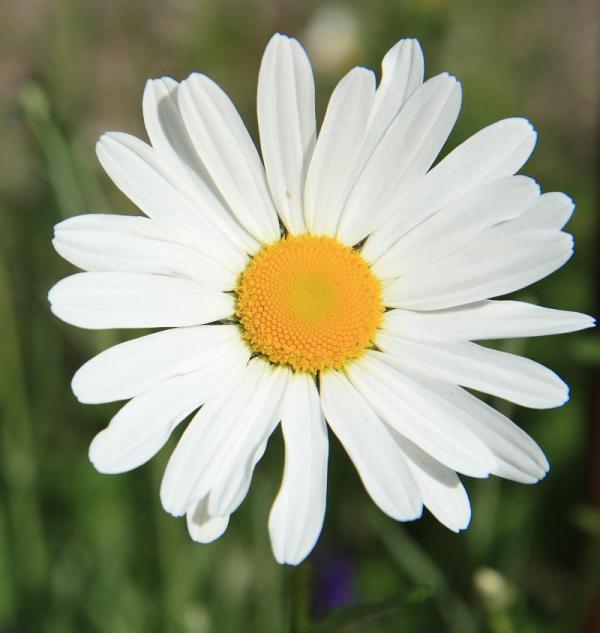}}

\begin{document}

\newcommand{\dd}{deg$^{2}$}
\newcommand{\flux}{$\rm erg \, s^{-1} \, cm^{-2}$}
\newcommand{\LL}{$\lambda$}

\vspace*{4cm}
\title{COSMOLOGY WITH X-RAY GALAXY CLUSTER SURVEYS ?}

\author{ Marguerite PIERRE}

\address{AIM, CEA, CNRS, Université Paris-Saclay, \\
Université Paris Diderot, Sorbonne Paris Cité, \\
F-91191 Gif-sur-Yvette, France
\vspace{0.5cm}
}

\maketitle\abstracts{
This talk reviews the scientific motivations, the potential difficulty and recent advances in cosmology using cluster number-counts in the  X-ray band. Our forward modelling approach shows that many of the practical and  conceptual shortcomings can now be overcome. We present recent results from the XMM-XXL survey.
\\
The next step is to apply artificial intelligence techniques on simulations. This allows us to bypass the unnecessarily complicated scaling relation formalism. The net gain is to significantly reduce the number of free parameters  and to provide  direct  access both to the cosmological parameters and  to truly physical ingredients, such as AGN feedback.\\
\underline{In this way, we achieve cluster cosmology without  explicit cluster mass calculation.} }

\section{A few historical landmarks}

The Uhuru mission undertook the first survey of the X-ray sky in 1970.  After 429 days of observations, 339 sources were discovered.  Out of these sources, 45 were associated with galaxy clusters \cite{giacconi74,forman78};  the angular resolution of the Uhuru collector was 30 arcmin. Subsequently, the Einstein observatory, the first imaging X-ray telescope (1978, 1 arcmin and 2 arcsec resolution) enabled the first study of X-ray emitting gas in clusters. Furthermore, the Einstein Medium Sensitivity Survey detected 733 serendipitous sources in the field of pointed targets, including 98 clusters;  this allowed for the attempt  to track back cluster evolution in terms of physical properties or number density  \cite{gioia90a,gioia90b}.  \\
The 1980s have been an exciting period in many respects for cosmology and we may highlight a few cornerstones, that nowadays belong to our obvious working landscape. Our view of the distant universe suddenly became much more structured with the publication of the CfA `Slice of the Universe', revealing that the universe is `bubbly' on very large scales \cite{lapparent86}. This motivated the development of topological tools beyond the 2-pt correlation function, like  the {\em genus curve} to decide whether the matter distribution is of cellular-, meatball- or sponge-like type \cite{gott88}. Numerical simulations, in the new CDM paradigm, were ramping up (\cite{davis85};
using 32768 particles, it collected 2850 citations!)
 while, in parallel, analytical calculations based on the model equation of non-linear diffusion (Burgers equation) achieved remarkable results \cite{gurbatov89}. Importantly, it was also realised that, under the hypothesis that only gravitation is at work, one can analytically predict the properties of the local cluster population and how this population evolves \cite{kaiser86}. There is only one physical scale in the problem : the mass-scale which is becoming non-linear at a given redshift. This yields simple scaling relations, a concept that subsequently constituted the skeleton of all  cluster cosmological studies.
Interestingly, C. Sarazin's book ({\em X-ray emission from clusters of galaxies}, 1988) which  was {\em the} reference cluster review by the ROSAT launch (1990), did not mention possible cosmological applications of clusters in his concluding remarks and future outlook. \\
The 1990s truly opened the era of cluster cosmological analyses. It was realised that cluster counts are degenerate with cluster evolution: at least, the knowledge of both dn/dz and of the evolution of the luminosity function is required \cite{oukbir97}. 
An analysis based on 70 ROSAT serendipitous clusters showed that their luminosity function does not evolve\footnote{which does not mean that clusters do not evolve} out to $z \sim 0.8$ \cite{rosati98}. Amazingly, this study that assumed a flux limited sample along with a cosmological model with $H_0=50$ and $q_0=1/2$, was corroborated 20 years later by the XMM-XXL cluster survey, in the framework of the WMAP9 cosmology \cite{pacaud16}! This is an interesting example of a `cosmological conspiracy' between cluster evolution and cosmology... An attempt to derive $\sigma_8$ and $\Omega_m$  using the evolution of the temperature function of a sample of 39 EMSS clusters, yielded constraints that are almost compatible with the Planck2018 CMB result at 1-sigma \cite{henry00}. A long term project aiming at inventorying and characterising  the clusters present in the Rosat All-Sky-Survey was undertaken by the REFLEX and NORAS teams: Constraints from the luminosity function involving some 900 southern clusters ($ 0<z<0.4$) showed a very good agreement with the then Planck cluster constraints  \cite{boehringer14}. The combination of the northern and southern samples (1653 clusters) revealed a local under-density in the universe, which would have a $\sim 5 \%$ effect on the measurement of $H_{o}$ \cite{boehringer20}. In the same direction, a 2021 study  handling five scaling relations for a sample of 570 clusters over the entire sky (also including XMM serendipitous clusters) revealed a $>5 \sigma$ anisotropy, hence suggesting a 9\% spatial variation of $H_o$ \cite{migkas21}.\\
The first XMM and Chandra observations confirmed that clusters are not simple objects; many physical processes are intricate and  questioned the use of clusters as cosmological probes. Over the past 20 years, considerable X-ray observing time  has been devoted  to deeply study single clusters in order to elucidate the processes at work in the intra-cluster medium. In the following, we focus on the complementary approach, based on blind cluster surveys.

\subsection{The XMM-XXL survey}
The XMM-XXL survey consists of two 25 \dd\ regions covered at a mean point source sensitivity of $ \sim 5\times 10^{-15}$ \flux\ in the [0.5-2] keV band  \cite{pierre16}. With 6.9 Ms total exposure time, XXL is currently the largest XMM programme and detected some 400 clusters out to a redshift of 2.
It is also the only XMM  `serendipitous' cluster survey (along with XCLASS, its companion project)  that provided results on $\sigma_8-\Omega_m$ so far. This fact underlines the difficulty of the task. Indeed, given the coverage and depth of the survey, several approximations used in the past needed to be revisited. Consequently the number of degrees of freedom in the analysis significantly increased. \\
In all that follows, we shall restrict the discussion to the cosmological analysis of cluster number counts.  Comprehensive reviews of cluster cosmology can be found  in  e.g. \cite{allen11,clerc22}.  
\subsection{Where do we stand now ?}
During the analysis of the XXL survey data, we unveiled and addressed a few important issues. Some 20 years later,   most of them are  taken for granted: \\
- Because clusters are extended objects, the cluster selection cannot be a simple flux limit, unless the limit is set well above the survey sensitivity (but then,  at the cost of the sample size). 
Moreover, the function must ideally depend solely on observable parameters; for  XXL it is the count rate and angular size of the sources. Subsequently,  the function can  be translated into the [M, z] space, for any cluster and cosmological model.
The selection function is not only critical for the cosmological analysis, but also for the determination of the scaling relations \cite{pacaud07,giles16}. \\
- The selection function  is tuned such as to favour `purity' against `completeness'. Indeed, the population of missed objects can be modelled for any cosmological+cluster physics model, while there is no model to a posteriori discard non-cluster or spurious detections.\\
- Scatter  in the mass-observable relations has a  critical effect on cluster counts. While increasing the scatter in the mass-temperature ($M-T$) or luminosity-temperature ($L-T$) relation increases the number of detected objects (similarly to $\sigma_8 $), increasing the scatter in the masse-core\_radius relation ($M-R_c$)  generally decreases the number of detected objects \cite{valotti18}.\\
- Consequently, scaling relations, scatters, selection function and cosmology must be fitted together: many free parameters!

\begin{figure}
\centerline{\includegraphics[width=9cm]{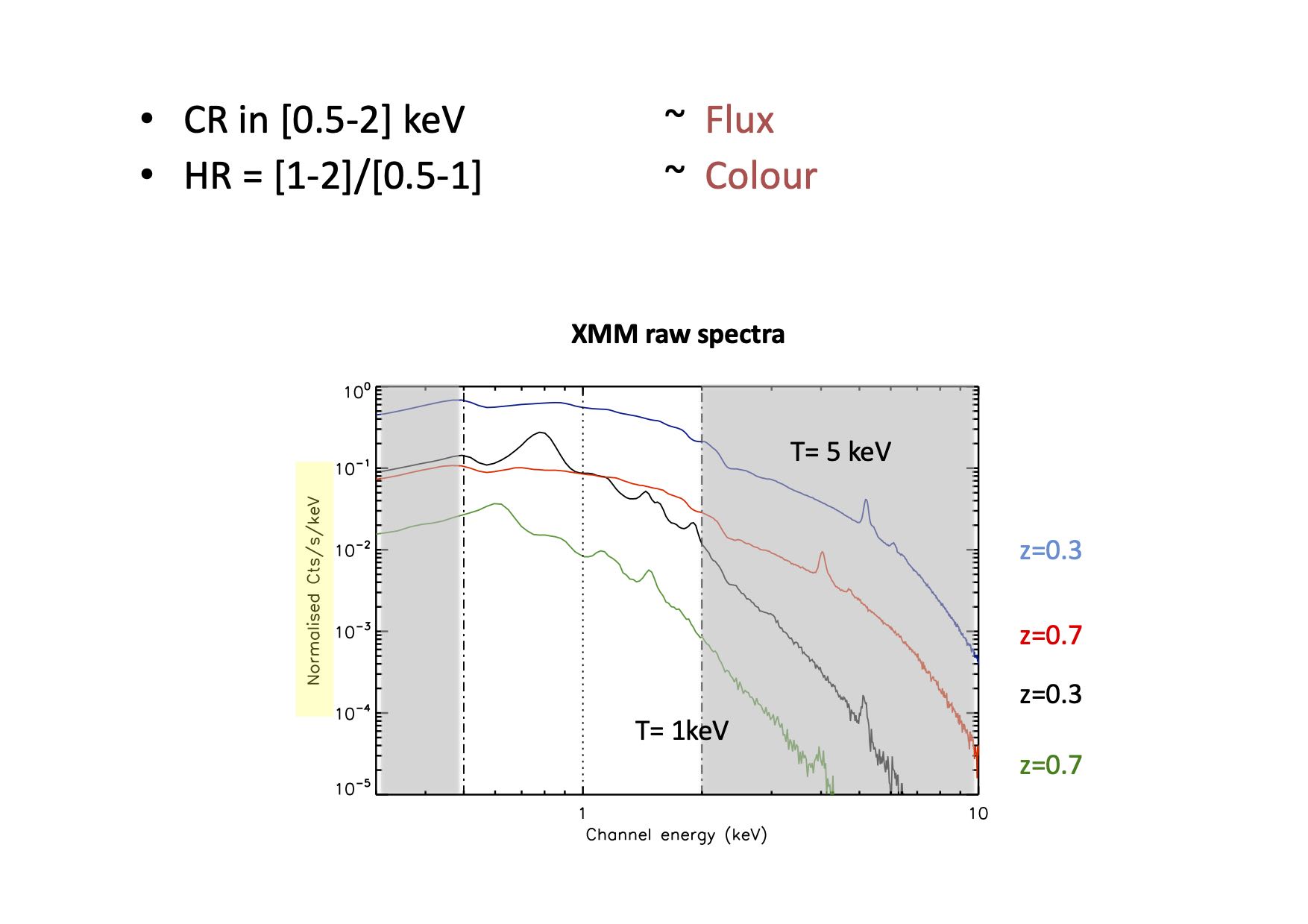} \hspace{-1cm}\includegraphics[width=9cm]{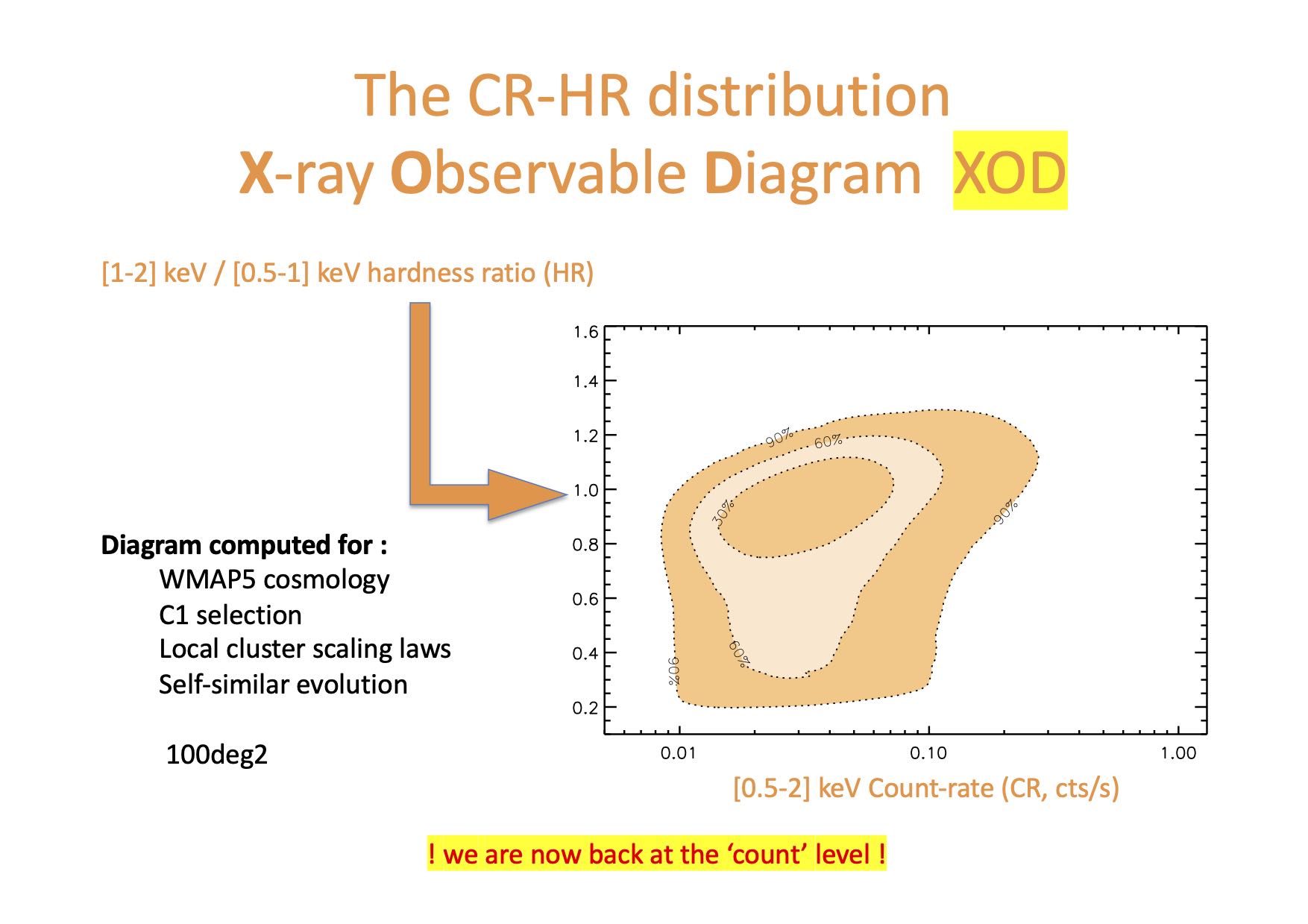}}
\caption[]{The  left panel shows different cluster spectra (for a heavy-element abundance of 0.3 solar) as seen by XMM. The 3 bands used to compute the count-rate (CR)  and hardness ratio (HR) are indicated in white. The Y-axis is in units of XMM counts.\\
The right panel shows the fiducial distribution of the CR and HR values for the cluster population detected  assuming the C1 XXL selection function and for the cosmology+physics set up indicated on the left. We note that the  presented  plot is integrated in the redshift dimension; for the cosmological analysis, the full 3-D X-ray Observable Diagram (XOD) is used.}
\end{figure}

\begin{figure}
\vspace{-0.5cm}
\centerline{\includegraphics[width=15cm]{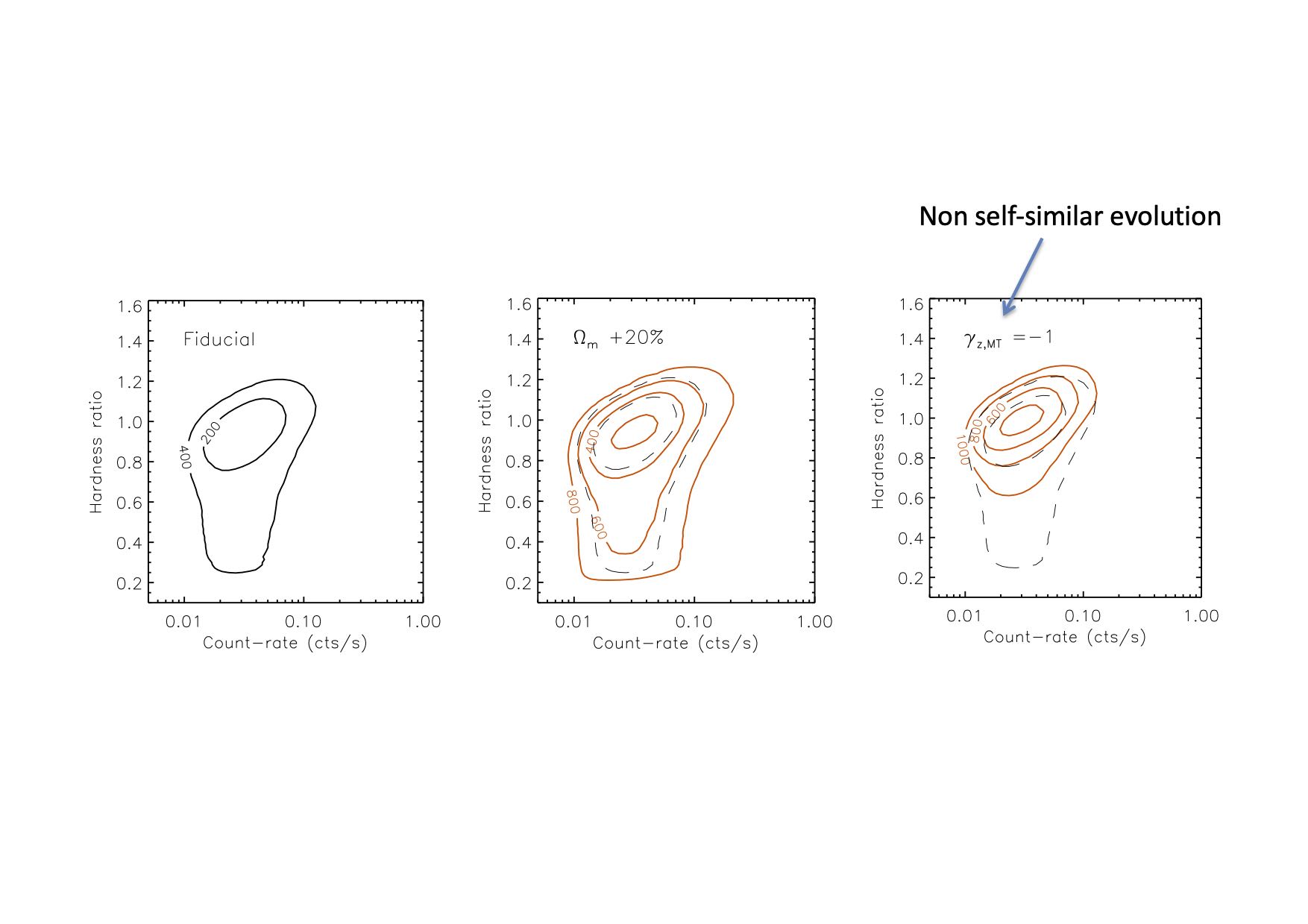}}
\vspace{-2cm}
\caption[]{Principle of the cosmological inference using XOD: effects of changing either the cosmological (middle) or the physics (right)  parameters   in the fiducial model (left) }
\label{fig:radish}
\end{figure}

\section{Forward cosmological modelling}
The implementation of the `traditional' method that maps cluster number counts into a mass distribution as a function of redshift requires heavy machinery. When running a MCMC analysis to iterate toward the most likely set of cosmological parameters,  the scaling relations must be recomputed for any considered cosmology - and this implies  recomputing as well the $R_{500}$ radius (and any other scaling parameter) in which fluxes and temperatures are integrated. Taking the problem the other way round, we developed a forward modelling approach that considers the simplest possible observed quantities, namely the XMM count-rates in three different bands (\cite{clerc12a,pierre17a}; ASpiX method).
Practically, a count-rate (CR) is analogous to a flux.  The ratio of two CRs (hardness ratio, HR) is equivalent to a colour and carries information about the cluster temperature and redshift. From this,  we construct `color-magnitude-redshift' diagrams of the detected X-ray cluster populations (XOD, see. Fig.1 \& 2). Fisher analyses demonstrated that the information contained in the XOD is at least as constraining for cosmology as the recovered cluster mass distribution as a function of redshift \cite{clerc12a}.  Because the HR determination requires much less photons than measuring cluster temperatures, the ASpiX method  allows the inclusion of all detected clusters down to 80-100 photons; this represents a considerable advantage compared to the `traditional' approach.

  \begin{figure}
\vspace{-1cm}
\centerline{\includegraphics[width=10cm]{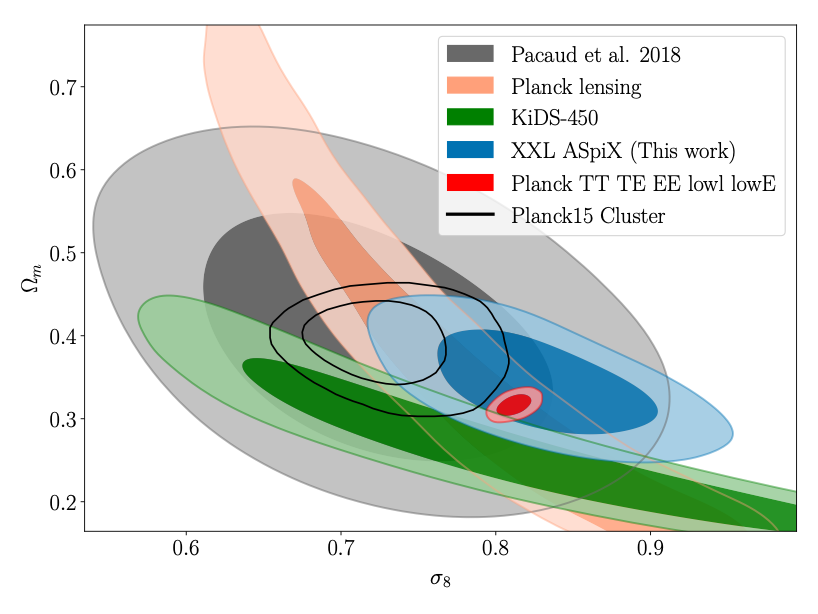}}
\caption[]{Comparison of the cosmological constraints from the 178 C1 XXL clusters, derived using either the dn/dz quantity (grey) or the CR-HR-z XOD (blue); both use the same priors.  Other surveys are   indicated: Planck CMB (red); Planck S-Z clusters (black  contours, 439 clusters); Planck lensing (orange); KIDS lensing (green). More details in \cite{garrel22a}}
\end{figure}

\subsection{Cosmological results}
In Fig. 3 we show recent results from the XXL cosmological analysis: constraints on $\sigma_8$ and $\Omega_m$   from cluster counts as a function of redshift are  compared to the outcome of the XOD analysis  from the same population (178 clusters) and  with the same degrees of freedom and priors. We obtain a factor of $\sim 2$ improvement on the cosmological constraints, assuming flat $ \Lambda$CDM \cite{garrel22a}.  In the same paper, the  impact of relaxing all scaling relation coefficients along with the implementation of  priors dynamically adapted to the cosmological model is also discussed. At the time of writing, calculations  related to the dark energy equation of state are being finalised (Garrel et al 2022b in prep).

\subsection{ A projection into the Athena era}
 While the XMM, eRosita and Euclid surveys essentially map the $0<z<1$ range,  the $1<z<2$ cluster universe will be systematically explored by Athena, the up-coming ESA X-ray mission  (launch by the mid-2030s, https://sci.esa.int/web/athena). The predicted redshift distribution of clusters to be detected by the Athena Wide Field Imager is at least one order of magnitude beyond the current surveys (Fig. 4): for the first time, we shall be in the position to accurately determine the X-ray cluster density above $z>1.2$ . We anticipate that a number of cosmological questions will remain open or  emerge after Euclid-eRosita.  Hence, Athena surveys will provide deep insight into  both cosmology  and cluster evolutionary physics, including the impact of AGN on the ICM, which is expected to be particularly efficient above $z>1$. A paper investigating these issues will be presented in the up-coming Athena `Red Book' (Cerardi et al in prep).

\begin{figure}
\vspace{-5cm}
\centerline{\includegraphics[width=9cm]{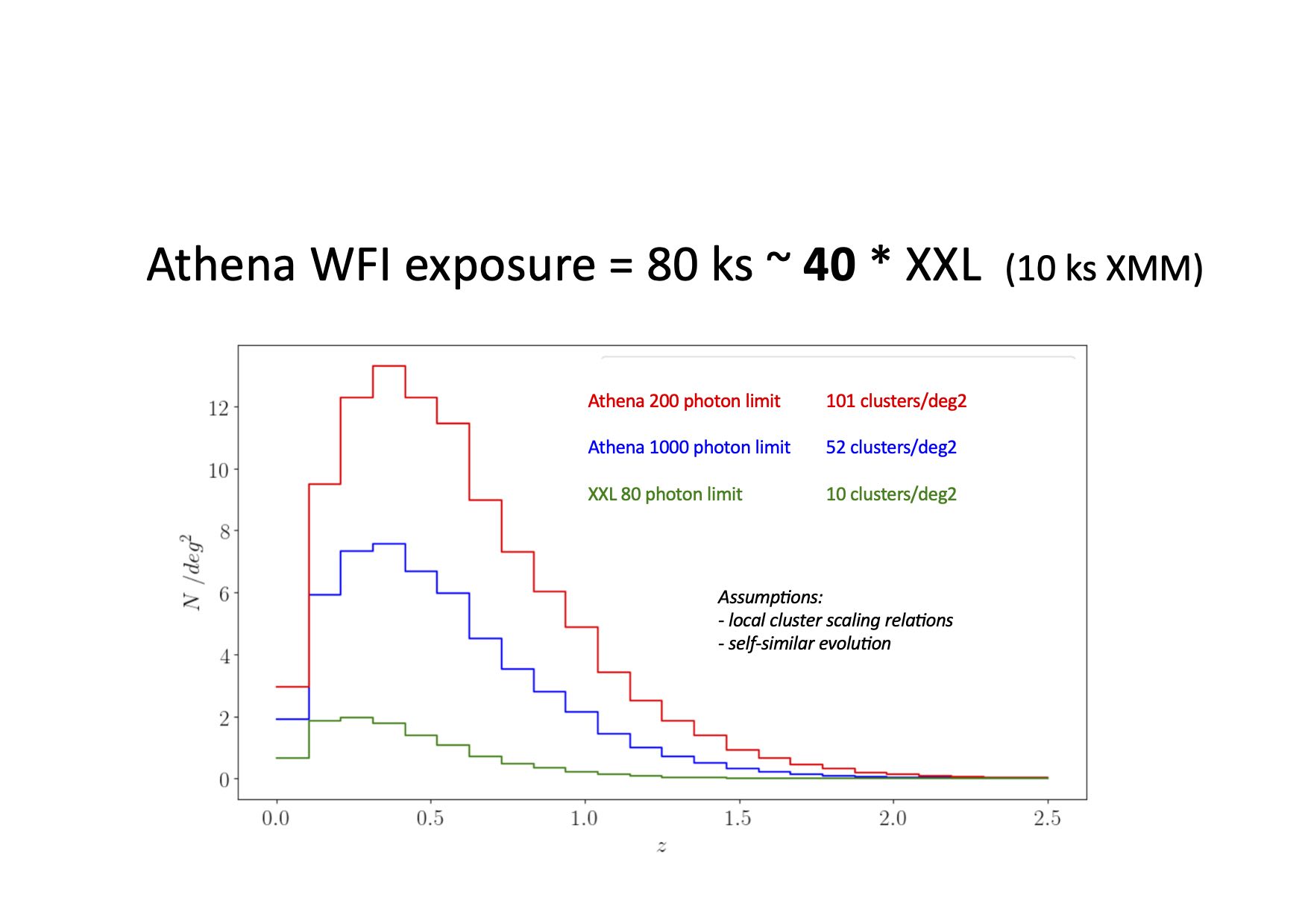}  \includegraphics[width=9cm]{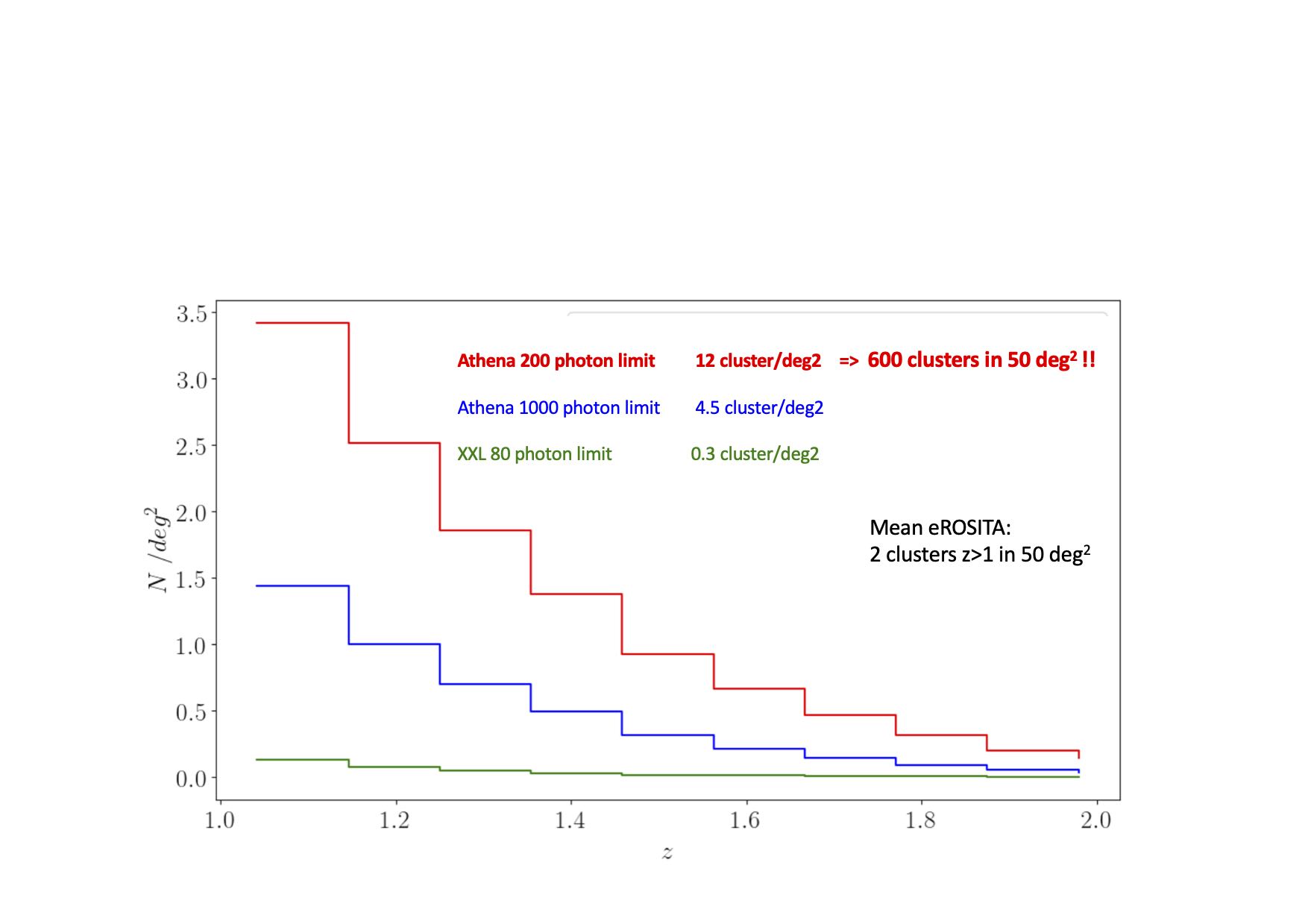}}
\caption[]{Predicted cluster number density to be detected by the Athena Wide Field Imager, assuming an exposure time of 80ks. This sensitivity is roughly  40 times deeper than the XMM-XXL survey. Two detection limits are considered (blue and red); for comparison, the XXL survey is shown in green. Athena will allow  in-depth cluster studies in the, so far poorly explored, $1<z<2$ range.
 [NB these estimates were performed assuming the current provisional Athena instrumental response - more detailed calculations will be presented in the Athena Red Book.]  }
\end{figure}

\section{The next future: combing artificial intelligence, numerical simulations and forward modelling}
While the ASpiX method marked a significant improvement  in the cosmological analysis of X-ray clusters, the current implementation is still burdened by two main factors: (1) the CR, HR translation to the mass function is implemented via scaling relations (even though cluster masses are not individually computed); (2) the MCMC analysis of the XOD is time-consuming (some 3 weeks with 5$\times$20 CPUs for the analysis presented in Fig.3). We are currently implementing Artificial Intelligence techniques to radically overcome these two hurdles. 
 
\subsection{Replacing the MCMC by a likelihood free inference method}
In this approach (e.g. \cite{charnock18}),
a neural network  compresses  the information contained in the XOD. The network is trained for a given cosmology over a  set of simulated XODs ($10^5$) covering a large range of random scaling relation coefficients. The dimension of the output vector  is the number of free cosmological parameters, while the training considers all scaling relation realisations as nuisance parameters.  This is rendered possible thanks to the 3-dimensional XOD representation that alleviates the degeneracy between cluster evolution and cosmology. The final cosmological inference corresponding to the observed XOD can subsequently be performed  by e.g.  an Approximate Bayesian Computation approach (Kosiba et al in prep).

\subsection{Overcoming the scaling relation formalism} 
Each scaling relation ($M-T,~ L-T,~ M-R_c$)  involves 4 parameters, i.e. slope, normalisation, evolution, scatter plus possibly  scatter evolution as well. This adds up to at least 12 free coefficients in addition to the cosmological parameters. However,  numerical simulations suggest that very few physical ingredients are actually needed to reproduce the bulk properties of the X-ray cluster population \cite{lebrun14}. It thus appears that the scaling relation formalism complicates the problem by  unphysically increasing the number of degrees of freedom and handling complex covariance matrices: many coefficient combinations are clearly degenerate over the mass range considered and for a given level of shot noise in the XOD (determined by the size of the surveyed area). It thus appears much more sensible to draw the XODs needed for step [3.1]  directly from hydrodynamic simulations that have been transformed into XMM images.  However, the computing time to produce a training sample of lightcones for a wide range of  cosmological and AGN feedback models  would be prohibitive.  We thus implemented a Larangian Deep Learning approach \cite[LDL]{dai20} to produce the required simulations. The model uses layers of Lagrangian displacements of
particles describing the observables to learn the effective physical laws; we train it on the CAMELS sample \cite{villaescusa21}. 

\subsection{Summary}
The final net outcome of the [3.2]+[3.1] procedure, when applied to an observed XOD, returns  the constraints on the cosmological parameters plus one or two  physics parameters (Cerardi et al in prep). The methodology totally bypasses any mass determination and derives quantities that have a direct physical interpretation. Clearly, a few more parameters remain more or less hidden in the course of the procedure, in particular in the simulations  (like the prescription for the black hole seeds and the resolution of the hydrodynamic solver). Next step will be to examine how these contribute to the final systematic error budget. \\
An obvious extension of the current work, would be to apply the ASpiX method to Sunyaev-Zel'dovich cluster samples, since the detection also relies on the intra-cluster medium properties. A significant difference, however, resides in the fact that the X-ray spectral information (the energy of the individual X-ray photons)  is a key ingredient of the XOD, while the S-Z spectral bands can only be used to assess the significance of the cluster detections; consequently,  in addition to the S-Z decrement, other direct observable quantities should be investigated like, for instance, apparent cluster sizes \cite{valotti18}, but this measure is usually very noisy.\\

{\em Acknowledgements:} We thank S. Madden, N. Cerardi, N. Clerc, C. Garrel, M. Kosiba,  J. B. Melin, F. Pacaud and P. Valageas for useful comments on the manuscript.

\vspace{2cm}

\footnotesize


\end{document}